\def\simgr{\mathbin{\;\raise1pt\hbox{$>$}\kern-8pt\lower3pt\hbox{$\sim$}\;}}
\def\simlr{\mathbin{\;\raise1pt\hbox{$<$}\kern-8pt\lower3pt\hbox{$\sim$}\;}}
\documentclass{aa}
\usepackage{epsfig}
\usepackage{amssymb}

\begin{document}
\thesaurus{06(08.18.1, 08.05.3, 08.13.2} 
\title{Stellar evolution with rotation VI: The Eddington and 
$\Omega$--limits, the rotational mass loss for OB and LBV stars}

\author{Andr\'e Maeder \& Georges Meynet}
   
\offprints{A. Maeder}

   \institute{Geneva Observatory,
              University of Geneva, 
              CH--1290 Sauverny,
	      Switzerland}
   \titlerunning{The $\Omega$-- and $\Gamma$--limits}
   \maketitle
   
   \markboth{A. Maeder and G. Meynet: Stellar Evolution with Rotation}
{A. Maeder and G. Meynet: Stellar Evolution with Rotation}

\begin{abstract}
Several  properties of massive stars with large effects of rotation 
and radiation are studied. For stars with shellular rotation, i.e. stars
with a constant  angular velocity $\Omega$ on horizontal surfaces
(cf. Zahn \cite{Za92}), we show that the equation of stellar surface has
no significant departures with respect to the Roche model;
high radiation pressure does not modify this property. Also, 
we note that 
contrarily to some current expressions, the correct Eddington factors
$\Gamma$  in a rotating star explicitely depend on
rotation. As a consequence, the maximum possible stellar luminosity 
is reduced by rotation.

We  show that there are 2 roots for the equation giving
the rotational velocities at break--up: 1) The usual solution,
which is shown to apply when the Eddington ratio $\Gamma$ of the star is
smaller than  formally 0.639. 2) Above this value of $\Gamma$, there 
is a second root, inferior to the first one, for the break--up velocity. 
This second solution 
tends to zero, when $\Gamma$ tends towards 1. 
This second root results from the interplay of radiation and rotation,
and in particular from the reduction by rotation of
the effective mass in the local Eddington factor.
The analysis made here should  hopefully 
clarify a recent debate between
Langer (\cite{La97,La98}) and Glatzel (\cite{Gla98}).
 
The expression for the global mass loss--rates 
is a function of
both  $\Omega$ and  $\Gamma$, and this 
may give raise
to extreme mass loss--rates ($\Omega \Gamma $--limit).
In particular, for O--type stars, LBV
stars, supergiants and Wolf--Rayet stars, 
even  slow rotation may dramatically enhance the mass loss rates.
Numerical examples  in the range of
9 to 120 M$_{\odot}$ at various  $T_\mathrm{eff}$  are given.

Mass loss  from rotating stars is anisotropic. 
Polar ejection is favoured by the  higher $T_\mathrm{eff}$ at the 
polar caps ($g_\mathrm{eff}$--effect),
while the ejection of an equatorial ring  is  favoured by the opacity
effect ($\kappa$--effect), if the opacity grows
 fastly for decreasing $T_\mathrm{eff}$.

\end{abstract}


\section{Introduction}

Recent models of stellar evolution with rotation (Meynet and Maeder
\cite{MM00}) have shown that rotation heavily modifies all the model 
outputs for massive stars. In the course of the above mentioned
work, it was realized that several basic points in the stellar physics
need to be further clarified, since they may have some
important consequences on the evolution.
 These points concern in particular the correct expression  of the
break--up velocities and  the dependence of 
the mass loss rates $\dot{M}$ on the observed rotation 
velocities $v$.  These problems are of great concern
for the most luminous stars close to the Eddington limit, like
OB stars,  supergiants, LBV and WR stars.

There is an interesting debate in recent litterature about what
is the correct expression for the critical velocity and what is
 the dependence of the  $\dot{M}$--rates on the rotation velocities
$v$. The critical rotation velocity of a star is often written
as $v^2_{\rm{crit}} = \frac{GM}{R} (1-\Gamma)$,
where $\Gamma = L/L_{\rm{Edd}}$ is the ratio of the 
stellar luminosity to the Eddington luminosity. With this expression,
Langer (\cite{La97},
\cite{La98}) suggests that ``no matter the rotation rate may be,
it (the star) will arrive at critical rotation well before 
$\Gamma$ = 1 is actually reached''. Consequently, Langer
introduces the concept that the stars generally reach the 
break--up limit, i.e. the $\Omega$--limit, earlier in evolution
than the $\Gamma$--limit.

This point of view was disputed by Glatzel(\cite{Gla98}), who 
stressed that the $\Omega$--limit is an artefact based on the
disregard of gravity darkening and on the assumption 
of a  uniform brightness over the surface of rotating
stars. Glatzel concludes that the Eddington 
factor has no effect on the critical rotation. This problem
needs to be further examined and this is what is done here.

Another important issue is the dependence of the mass loss rates
$\dot{M}$  on rotation velocities $v$ (cf. Maeder and Meynet \cite
{MaMe00}).
On one side, Langer (\cite{La97}, \cite{La98}), Heger et al. 
(\cite{He00}), Meynet and Maeder (\cite{MM00}) are using a 
relation $\dot{M}$ vs. $v$ from
Friend and Abbott (\cite{Fr86}) which formally leads to infinite
mass loss rates at break--up velocities. On the other side,
Owocki et al. (\cite{Ow96,Ow98}), Owocki and Gayley (\cite{Ow97}) and 
Glatzel (\cite{Gla98}) show that even at extreme rotation the mass
loss rates remain finite and do not differ too much from the case of 
zero--rotation. There also, some further analysis is needed.

In a rotating star, the total gravity is the sum of the gravitational, centrifugal and radiative accelerations:

\begin{equation}
 \vec{g_\mathrm{tot}} = \vec{g_\mathrm{eff}} + \vec{g_\mathrm{rad}} = \vec{g_\mathrm{grav}} +
 \vec{g_\mathrm{rot}} + \vec{g_\mathrm{rad}} .
\end{equation}

\noindent
For purpose of clarity, we adopt the following definitions:

\vspace{2mm}
\noindent
-- We speak of the Eddington or $\Gamma$--limit,
when rotation effects can be neglected  and 
$\vec{g_\mathrm{rad}} + \vec{g_\mathrm{grav}} = \vec{0}$, 
which implies that

\begin{equation}
\Gamma = \frac{\kappa L}{4 \pi c GM}  \; \rightarrow  \; 1.
\end{equation}

\noindent In that case 
$L = L_{\rm EDD}=4 \pi c GM/\kappa$.
The opacity $\kappa$ considered here is the total opacity, unless we
specify it differently (cf. Sect. 4.2).

\vspace{2mm}
\noindent
--The break--up or $\Omega$--limit is reached, for a star
with an angular velocity $\Omega$ at the surface, when the effective
gravity  $\vec{g_\mathrm{eff}} = \vec{g_\mathrm{grav}} +
 \vec{g_\mathrm{rot}} = \vec{0}$ and in addition
when radiation pressure effects can be neglected.

\vspace{2mm}
\noindent
--\textit{The $\Omega \Gamma$--limit is  reached when the total gravity
$\vec{g_\mathrm{tot}} = \vec{0}$, with significant effects of both
rotation and radiation.} 
This is the  general case, that we study here. It should
lead to the two above cases in their respective limits.

\vspace{2mm}
In Sect. 2, we examine the 
surface gravity, the Eddington factors and the limiting
luminosity.  In Sect. 3, the expression of the break--up
velocities are considered, while in Sect. 4 we examine the 
mass loss rates. The equation of the surface
with account of rotation and radiative acceleration is discussed in an
Appendix.

\section{Surface gravity, Eddington factors and limiting luminosity}

\subsection{The von Zeipel Theorem close to the 
$\Omega \Gamma$--limit}

The von Zeipel theorem (\cite{vZ24})  expresses that the radiative
 flux $\vec{F}$
at some colatitude $\vartheta$ in a rotating star is proportional 
to the local effective  gravity $\vec{g_\mathrm{eff}}$.
 In a previous work (Maeder \cite{Ma99}),
we have generalized this theorem to the case of shellular rotation
proposed by Zahn (\cite{Za92}). Shellular rotation
results from strong horizontal
turbulence which reduces the latitudinal dependence of rotation and
makes the angular velocity $\Omega$ constant on  an isobar.
Here, we shall consider the case of stars with shellular rotation,
where the $\Omega \Gamma$--limit may play a role. 

As shown by Langer (\cite{La97}), stars close to the Eddington limit 
tend to develop convection in the outer
layers (cf. also Maeder \cite{Ma80}). However, in the outer layers the
convective flux is generally negligible and the main transport mechanism is
radiative transfer, a point also emphasized  by Glatzel (\cite{Gla98}).
As a numerical example, in a 60 M$_{\odot}$ model 
(Meynet and Maeder \cite{MM00}) at the end of the MS phase with
log L/L$_{\odot}$= 5.89 and  log T$_{\rm{eff}}$ =4.34 , 
the convective flux is  negligible  down to a fractional radius r/R
 of 0.85. In a 120 M$_{\odot}$ with log L/L$_{\odot}$ = 6.32 and 
 log T$_{\rm{eff}}$ =4.35 at the end of the
MS, the convective flux is negligible  down to r/R = 0.63.
Thus, the basic condition to apply the von Zeipel theorem
to stars close to the  $\Omega \Gamma$--limit is fulfilled, since the flux
is  essentially radiative. 

The  expression of the flux $\vec{F}$ (Maeder \cite{Ma99}) for a star with
angular velocity $\Omega$ on the isobaric stellar surface (cf.
Appendix) is

\begin{equation}
\vec{F}  =  - \frac{L(P)}{4 \pi GM_{\star}}
\vec{g_{\rm{eff}}} [1 + \zeta(\vartheta)]  \quad {\mathrm{with}} \quad 
\end{equation} 
 
\begin{equation}
M_{\star} = M \left( 1 - \frac{\Omega^2}
{2 \pi G \rho_{\rm{m}}}  \right)
\quad {\mathrm{and}} \quad
\end{equation}

 \begin{equation}
\zeta(\vartheta) = \left[\left(1 - \frac{\chi_T}{\delta}\right) \Theta +
\frac{H_T}{\delta} \frac{d\Theta}{dr}\right] P_{2}(\cos \vartheta).
\end{equation}

\noindent
There, $\rho_{\rm{m}}$ is the internal average density,
 $\chi = 4acT^3/(3 \kappa \rho)$ and $\chi_{T}$
is the partial derivative with respect to $T$.  The quantity 
$\Theta$ is defined by 
$\Theta = \frac{\tilde{\rho}}{\bar{\rho}}$, 
i.e. the ratio of the horizontal density fluctuation
to the average density on the isobar, which is given by
$ \frac{\tilde{\rho}}{\bar{\rho}} = \frac{1}{3} \frac{r^2}{\bar{g}}
\frac{d\Omega^{2}}{dr}$ where $\bar{g}$ is the average gravity
on an isobar (cf. Zahn \cite{Za92}).
One has the thermodynamic coefficients $\delta = - (\partial \ln\rho /
 \partial \ln T)_{P, \mu}$, $H_{T}$ is the temperature scale height.
 The term $\zeta(\vartheta)$, 
which expresses the deviations of the von Zeipel theorem due to the 
baroclinicity of the star,
is generally very small (cf. Maeder \cite{Ma99}).

Let us emphasize that the flux is proportional to $\vec{g_\mathrm{eff}}$
and not to $\vec{g_\mathrm{tot}}$. This results from the fact that
the equation of hydrostatic equilibrium is 
$\frac{\vec{\nabla} P}{\rho} = - \vec{g_\mathrm{eff}}$.
The effect
of radiation pressure is already counted in the 
expression of $P$, which is the total pressure.
We may call $M_{\star}$ the effective mass, i.e. the mass
reduced by  the centrifugal
force. This is the complete form of the von Zeipel theorem in a 
differentially rotating star with shellular rotation, whether or not 
one is close to the Eddington limit.

\subsection{Expressions of the gravity and of the 
local Eddington factor}

Let us express  the total gravity at some colatitude 
$\vartheta$, taking into account the
radiative acceleration

\begin{equation}
 \vec{g_\mathrm{rad}} = \frac{1}{\rho} \vec{\nabla} P_\mathrm{rad} = \frac{\kappa(\vartheta)\vec{F}}{c} \; ,
\end{equation}

\noindent
thus one has with Eq. (1.1), (2.3) and (2.4)

\begin{eqnarray}
\vec{g_\mathrm{tot}}    =
 \vec{g_\mathrm{eff}} \left[1 -  \frac{\kappa(\vartheta)L(P) 
[1 + \zeta (\vartheta)] }
{4 \pi cGM (1 - \frac{\Omega^2} {2 \pi G \rho_{\mathrm{m}} } )}\right].
\end{eqnarray}

\noindent
The rotation effects appear both in $\vec{g_\mathrm{eff}}$ and in the 
term in brackets. When we write $\kappa(\vartheta)$, we mean 
that in a rotating star, the local T$_\mathrm{eff}$ and gravity
vary with latitude and so does the opacity.We may also consider the local limiting flux.
The condition $\vec{g_\mathrm{tot}}= \vec{0}$ in Eq. (1.1)
 with  Eq. (2.6) for $\vec{g_\mathrm{rad}}$ allows us to
define a limiting flux,

\begin{equation}
\vec{F_{\mathrm{lim}}}(\vartheta) = - \frac{c}{\kappa(\vartheta)}  
\vec{g_{\mathrm{eff}}}(\vartheta) \; .
\end{equation}

\noindent
From that we may define the ratio  $\Gamma_{\Omega}(\vartheta)$
of the  actual flux 
$F(\vartheta)$ to the limiting local flux in a rotating star,

\begin{eqnarray}
\Gamma_{\Omega}(\vartheta) =
\frac{F(\vartheta)}{F_{\mathrm{lim}}(\vartheta)}=
\frac{ \kappa (\vartheta) \; L(P)[1+\zeta (\vartheta)]}{4 \pi 
cGM \left( 1 - \frac{\Omega^2}{2 \pi G \rho_{\rm{m}}}  \right) } \; .
\end{eqnarray}

\noindent
 As a matter of fact,
 $\Gamma_{\Omega}(\vartheta)$ is the local Eddington ratio.
For zero rotation $\Gamma_{\Omega}(\vartheta) = \Gamma$ as
given by Eq. (1.2).
Using relation (2.9), we may
write the Eq. (2.7) for the total gravity  as

\begin{equation}
\vec{g_\mathrm{tot}} = \vec{g_\mathrm{eff}}
\left[ 1 - \Gamma_{\Omega}(\vartheta) \right] \; .
\end{equation}

\noindent
This  shows that the 
expression for the total acceleration in a rotating star is 
similar to the usual one, except  that $\Gamma$ is 
replaced by the local value  $\Gamma_{\Omega}(\vartheta)$. 
Indeed, contrarily to expressions such as $\vec{g_\mathrm{tot}} = \vec{g_\mathrm{eff}} \left( 1 - \Gamma \right)$
often found in literature, we see that the appropriate Eddington factor
(2.9) also depends on the angular velocity $\Omega$ on the 
isobaric surface. 

From (2.9), we note that over the surface of a rotating star, which
has a varying gravity and T$_\mathrm{eff}$,  $\Gamma_{\Omega}(\vartheta)$
is the highest at the latitude where $\kappa(\vartheta)$ is the largest,
(if we neglect the effects of $\zeta (\vartheta)$, which is justified 
in general). If the opacity increases with decreasing T as in  hot
stars, the opacity is the highest at the equator and there the limit
 $\Gamma_{\Omega}(\vartheta) = 1 $ may be reached first. Thus, it is
to be stressed that if the limit  $\Gamma_{\Omega}(\vartheta) = 1 $ 
happens to be met at the equator, it is not because 
$\vec{g_\mathrm{eff}}$ is the lowest there, but because the 
opacity is the highest ! Indeed, both dependences in $\vec{g_\mathrm{eff}}$
have cancelled each other in the ratio given by Eq. (2.9).

\subsection{The  luminosity at the $\Omega\Gamma$--limit}

The $\Omega \Gamma$--limit is reached, when 
the  local Eddington ratio  $\Gamma_{\Omega}(\vartheta) = 1$ 
at some colatitude $\vartheta$. The condition 
$\Gamma_{\Omega}(\vartheta) = 1$ allows us to define  a limiting 
luminosity $L_{\Omega \Gamma}$ at the
$\Omega \Gamma$--limit, i.e. when both the effects of radiative 
acceleration and rotation are important. From (2.9) we have

\begin{equation}
L_{\Omega \Gamma} = \frac{4 \pi c G M}{\kappa(\vartheta)
\left[1+\zeta (\vartheta)\right]}
\left( 1 - \frac{\Omega^2}
{2 \pi G \rho_{\rm{m}}} \right)  \; .
\end{equation}

\noindent
It means that for a certain angular velocity $\Omega$ on the 
isobaric surface, the maximum permitted luminosity of a star is
reduced by rotation, with respect to the usual Eddington limit
(cf. Sect. 1).
This conclusion was also reached by Glatzel (\cite{Gla98}).
In the above relation, $\kappa(\vartheta)$ is the largest value of
the opacity on the surface of the rotating star.
For O--type stars with photospheric opacities dominated
by electron scattering, the opacity $\kappa$ is the same
everywhere on the star.
 For the equation of the surface discussed in the Appendix,
the maximum value of $\frac{\Omega^2} {2 \pi G \rho_{\rm{m}}} = 0.361$,
(with more digits it is 0.360747).

\section{The break--up velocities}

We have seen above that rotation may be considered as reducing the
maximum possible luminosity for a star. An alternative way
to consider the problem is to ask
the question: what happens to the break--up velocity for a star close
to the Eddington limit ? Most authors (Langer \cite{La97, La98, La99};
Lamers et al. \cite{Lam99}; Heger et al. \cite{He00}) write 
this critical velocity $v_{\rm{crit}}$ like

\begin{equation}
v^2_{\rm{crit}} = \frac{GM}{R} (1-\Gamma) \; .
\end{equation}

\noindent
This relation is true if we assume that the brightness of
the rotating star is uniform over its surface, which is in
contradiction with von Zeipel's theorem. Surprisingly, some 
authors use this relation simultaneously with the von Zeipel theorem.
Eq. (3.12), which we do not support, in agreement with
 Glatzel (\cite{Gla98}), implies that
the break--up velocity is reduced by the proximity to the
Eddington limit.  The problem needs to 
be  further studied carefully.

The critical velocity is reached when somewhere on the star one has
$\vec{g_\mathrm{tot}}=\vec{0}$, i.e. according to (2.10)

\begin{equation}
 \vec{g_\mathrm{eff}} \;\left[1 - \Gamma_{\Omega}(\vartheta)
 \right] = \vec{0} .
\end{equation}

\noindent
This equation has two roots. The first one $v_\mathrm{crit,1}$
is given by the usual condition
$\vec{g_\mathrm{eff}}= \vec{0}$, which implies the equality
$\Omega^2 R^{3}_\mathrm{eb}/(GM) =1$ at the equator (cf. Eq. A2
in Appendix). This corresponds to an equatorial critical velocity

\begin{equation}
v_\mathrm{crit, 1} = \Omega \; R_\mathrm{eb} = 
\left( \frac{2}{3} \frac{GM}{R_\mathrm{pb}} \right)^{\frac{1}{2}} \; .
\end{equation}

\noindent
$R_\mathrm{eb}$  and $R_\mathrm{pb}$ are respectively the 
equatorial  and polar radius at  the break--up velocity and they obey to
the surface equation. We notice
that the critical velocity $v_\mathrm{crit, 1}$ is independent
on the Eddington factor. To this extent, this is in agreement with
Glatzel (\cite{Gla98}). The basic physical reason for this
independence is quite clear:
the radiative flux decreases at the equator, when the effective gravity
decreases.

Equation (3.13)  has a second root, which is given by the condition
$\Gamma_{\Omega}(\vartheta)$ = 1. As seen above, this condition
will in general be met at the equator first.
We thus have to search for
the corresponding critical velocity $v_\mathrm{crit, 2}$ for a given 
value of the stellar luminosity. This second root  has to be compared
to the first one. For given values of M and  L,
the lowest of the two roots $v_\mathrm{crit, 1}$ and $v_\mathrm{crit, 2}$
is the significant one, since as soon as it will be  reached 
the matter at the surface of the star is no longer bound. The condition
$\Gamma_{\Omega}(\vartheta)$ = 1 gives, if we neglect $\zeta (\vartheta)$,

\begin{equation}
\frac{\kappa (\vartheta) L(P)}{4 \pi cGM} = 1 - \frac{\Omega^2}
{2 \pi G \rho_{\mathrm{m}}}.
\end{equation}

\noindent
Let us write

\begin{equation}
\frac{\Omega^2}{2 \pi G \rho_{\mathrm{m}}} \;= 
 \; \frac{16}{81} \omega^2 V^{\prime}(\omega) 
\quad {\mathrm{with}} \quad
\end{equation}

\begin{equation}
V^{\prime}(\omega)= \frac{V(\omega)}{\frac{4}{3} \pi R^3_\mathrm{pb}} 
\quad {\mathrm{and}} \quad
\omega^2 = \frac{\Omega^2 R^{3}_\mathrm{eb}}{GM} .
\end{equation}

\noindent
The quantity $\omega$ is the fraction 
 of the angular velocity at the classical  break--up  given by Eq. (3.14).
The density $\rho_\mathrm{m}(\omega) = M/V(\omega)$, 
where the stellar volume  $V(\omega)$ depends on rotation. The quantity $V^{\prime}(\omega)$
is the ratio of the actual volume of a star with rotation $\omega$ to 
the volume of a sphere  of radius   $R_\mathrm{pb}$.
$V^{\prime}(\omega)$ is  obtained by the integration of the
solutions of the surface equation (A6) for a given value of
the parameter $\omega$. At break--up velocity  $v_\mathrm{crit, 1}$ , the value of 
$V^{\prime}(\omega)$ = 1.829, which gives the maximum value of
$\frac{\Omega^2}{2 \pi G \rho_{\mathrm{m}}}$ = 0.361.
If we call $\Gamma_\mathrm{max}$ the maximum Eddington ratio
$\kappa(\vartheta) L(P)/(4 \pi c GM)$ over the surface 
(in general at the equator), 
Eq. (3.15) can thus be written

\begin{equation}
\frac{16}{81} \;  \omega^2 \; V^{\prime}(\omega) \; = 
\; 1- \Gamma_\mathrm{max} \; .
\end{equation}

\noindent
For a given value of  $\Gamma_\mathrm{max}$,
 one must search the value of
$\omega$ which satisfies this equation.
 This is easily obtained by solving 
numerically the surface equation. 
For a given large enough
 $\Gamma_\mathrm{max}$ (i.e. larger than 0.639), 
the obtained $\omega$--value is  
lower than 1, and this implies a corresponding critical
velocity $v_\mathrm{crit, 2} \;$ given by

\begin{eqnarray}
v_\mathrm{crit, 2}^2 = \Omega^2 R^{2}_\mathrm{e}(\omega) =
\frac{81}{16} \; \frac{1-\Gamma_{\mathrm{max}}}{V^{\prime}(\omega)} 
\frac{GM}{R^3_{\mathrm{eb}}}
R^{2}_\mathrm{e}(\omega) = \nonumber \\[2mm]
\frac{9}{4} \;v_\mathrm{crit, 1}^2 \; 
\frac{1-\Gamma_{\mathrm{max}}}{V^{\prime}(\omega)} \; \frac{
R^{2}_\mathrm{e}(\omega)}{R^2_{\mathrm{pb}}} \; ,
\end{eqnarray}

\begin{figure}[tb]
  \resizebox{\hsize}{!}{\includegraphics{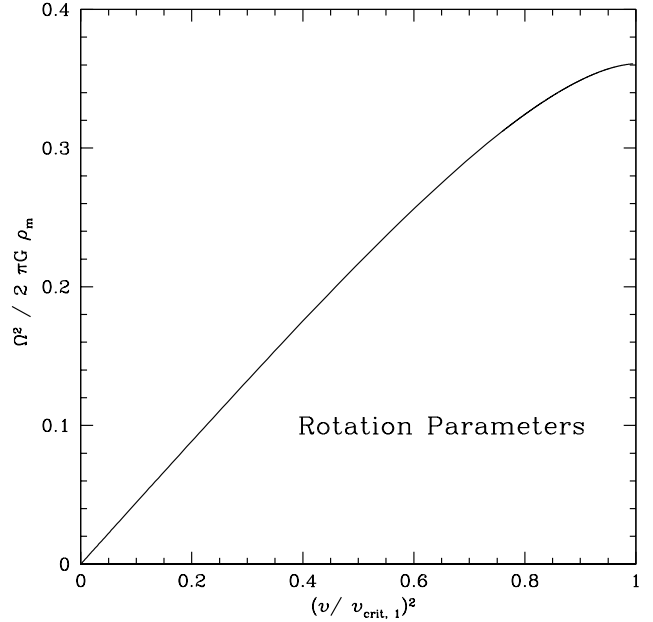}}
  \caption{The critical velocity $v_\mathrm{crit, 2}$ expressed as
a fraction of $v_\mathrm{crit, 1}$ plotted as a function
of the Eddington factor   $\Gamma_\mathrm{max}$, which is 
the largest value of the Eddington factor
reached where the opacity is the largest over
the stellar surface. We notice that when 
the Eddington factor tends towards unity, the critical velocity
goes down to zero.}
\label{rotgamma}
\end{figure}

\noindent
where R$_\mathrm{e}(\omega)$ is the equatorial radius for a given 
value of the rotation parameter $\omega$.
Fig.~\ref{rotgamma} illustrates the results. We notice  that the second 
root $v_\mathrm{crit, 2}$, expressed as a fraction of
the first root $v_\mathrm{crit, 1}$
given by (3.14),  tends towards zero when the Eddington factor 
$\Gamma_\mathrm{max}$
tends towards 1. The first root defined by (3.14) can still be
written, but the second root defined  by the condition 
$\Gamma_{\Omega}(\vartheta)=1$ is met first.
This reduction of the critical velocity with
respect to the classical expression only occurs for Eddington factors
larger than 0.639, since  the maximum value of $\frac{\Omega^2} {2 \pi G \rho_{\rm{m}}}$ is  0.361 (cf. Eq. 3.15).  

This second root results physically from  both effects of
rotation and radiation: for   $\Gamma_\mathrm{max} > 0.639$,
a zero value of $\vec{g_\mathrm{tot}}$ can be achieved for
non--extreme
rotations. This enters through the reduction due to rotation of
the effective mass M$_{\star}$,
which is the significant mass in the local Eddington factor
$\Gamma_{\Omega}(\vartheta)$.

For  $\Gamma_\mathrm{max} < 0.639$,
Eq. (3.18) has no solution. Physically this means that
when the star is sufficiently far from the Eddington limit, 
the reduction of the effective mass M$_{\star}$ by rotation
is not sufficient to bring $\Gamma_{\Omega}(\vartheta)$  to 1.
In that case, Eq. (3.13) has only 
one root given by the classical Eq. (3.14).

We hope that these results clarify the
debate between  Glatzel (\cite{Gla98})
and Langer(\cite{La97}). On one hand, we see that the claim by Langer that 
stars close to the Eddington limit have a lower rotation limit is correct,
even if the  Eq. (3.12) by Langer is not the right one. On the other
side, Glatzel has claimed that the Eddington factor does not affect
the break--up velocity,  we see that this is true in general for most stars
for the reasons given above, however for 
 $\Gamma_\mathrm{max} \geq 0.639$ this statement does not apply.

The moment when stars reach their critical velocities is
far from being an academic one, since when this occurs  large
mass loss enhancements  may result, a point which is examined below.

\section{The mass loss rates as a function of $\Omega$ and 
$\Gamma$}

\subsection{Present context}

The effects of rotation on the mass loss rates have
been studied both observationally and theoretically. Observationally, very
large changes of the $\dot{M}$--rates, i.e. up to
 2--3 orders of a magnitude,
were suggested by Vardya (\cite{Var85}). However, Nieuwenhuijzen and
de Jager (\cite{Nieu88}) claimed with reason
that the correlation found by Vardya was 
largely the reflect of the distribution of the mass loss rates
and rotational velocities over the HR diagram. When 
disentangling  the various effects, 
 Nieuwenhuijzen and de Jager found much smaller  effects
of rotation.  However, they noticed that the $\dot{M}$--rates of the
Be--stars are larger by about a factor 10$^2$. Since Be--stars
are fast rotating stars, we may wonder 
whether the effects of rotation on the mass loss rates are
really so negligible, as these last authors considered them.
Certainly further observational studies are also required.

On the theoretical side, Pauldrach et al. (\cite{Pau86}),
Friend and Abbott (\cite{Fr86}) find only a moderate increase of the $\dot{M}$--rates, of about
30 \% for $v =350 \;$ km/s.
 Friend and Abbott find an increase of the $\dot{M}$--rates
which can be fitted by the relation (Langer \cite{La98};
 Heger and  Langer \cite{He98})
\begin{equation}
\dot{M}(v)\; = \;\dot{M} (v = 0) 
\left(\frac{1}{1-\frac{v}
{v_{\rm crit}}}\right)^{\xi}
\end{equation} 
\noindent with $\xi$ = 0.43. This 
expression, often used in evolutionary models, is based on wind models
which do not account for the von Zeipel theorem. 
We notice that Eq. (4.20) diverges at break--up,
while as shown by Glatzel (\cite{Gla98}) and Owocki et al. (\cite{Ow96,
Ow98}),
the stellar mass loss rates should not diverge at the $\Omega$--limit,
(however see Sect. 4.3).

When a proper account of 
the gravity darkening is made, there are two main
terms contributing to the anisotropic mass loss rates from a 
rotating star 
(cf. Maeder \cite{Ma99}).  1) The ``g$_\mathrm{eff}$--effect''
which favours polar ejection,  since the polar caps of a rotating
star are hotter. 2) The ``opacity or $\kappa$--effect'', which  may favour 
an equatorial ejection, when the opacity is large enough at the equator
due to  the lower T$_{\mathrm{eff}}$.

In O--type stars, since opacity is due 
mainly to the T--independent  electron scattering,  
the g$_\mathrm{eff}$--effect is likely to  dominate,  raising 
a fast highly ionized polar wind. In B-- and later type stars, the
opacity effect may favour a dense equatorial wind and ring formation,
with low terminal velocities and low ionization. Recently 
Petrenz and Puls (\cite{PP00}) have constructed 2--D models of line driven
winds for rotating O--type stars. They found that the mass loss 
from hot star is essentially polar due to the $g_{\rm eff}$--effect.
Quantitatively their results differ very little from previous works
by Pauldrach et al. (\cite{Pau86}).

\subsection{Expression of the $\dot{M}$--rates as a 
function of $\Omega$ and $\Gamma$}

It is worth to further  examine the consequences of the
above results on
the dependence of the mass loss rates on rotation. 
According to the radiative wind theory (cf. Castor et al. 
\cite{Ca75}; Pauldrach
et al. 1986; Kudritzki et al. \cite{Ku89}; Puls et al. \cite{Pu96}), we may write
 the mass loss  fluxes  $\Delta\dot{M}/ \Delta \sigma$  by surface
elements $\Delta \sigma$

\begin{equation}
\frac{\Delta\dot{M}(\vartheta)}{\Delta \sigma}
 \simeq \left(k\alpha\right)^{1/\alpha} 
\left(\frac{1-\alpha}{\alpha}\right)^{\frac{1-\alpha}{\alpha}}
F(\vartheta)^{1/\alpha} g_{tot}^{1-\frac{1}{\alpha}} (\vartheta)
\end{equation}

\noindent
where $k$ and $\alpha$ are the force multiplier parameters.
At some temperatures, the ionisation equilibrium of the stellar
wind is changing abruptly and so does the opacity of the plasma.
Consequently, the values of the force multipliers undergo rapid 
transitions for certain values of $T_\mathrm{eff}$, particularly at
21 000 K and maybe also at 10 000 K (cf. Lamers et al. \cite{Lam95};
Lamers \cite{Lam97}). Such fast
transitions of the wind properties are  called by Lamers a 
bi--stability of the stellar winds, since near the transition
limit the wind can exist in two  states.
There are both empirical and theoretical determinations 
of $\alpha$, however they lead to rather different values
(cf. Lamers et al. \cite{Lam95}). The empirical ones, 
based on the values of the  observed terminal velocities, are
in general smaller than the theoretical estimates. As empirical values,
 Lamers et al. (\cite{Lam95}) obtain, for example,
 $\alpha$ = 0.52 for $ 4.70 \geq \log T_\mathrm{eff} \geq 4.35$, (type B1.5
or earlier);
$\alpha$ = 0.24, 0.21, 0.17, 0.15 for
 $\log T_\mathrm{eff}$ = 4.30 (type B2.5),
4.20 (B5), 4.00 (B9.5), 3.90 (A7) respectively. 
These transitions may produce jumps in the mass loss rates, with the
high rates on the low side of the transition. 
As  $T_\mathrm{eff}$ is decreasing from the pole
to the equator, one may thus expect, on the surface of a fast 
rotating star of
type B or later, the occurence of some bi--stability limits and
the corresponding variations of the force multipliers and of
the mass loss rates. In a star, where a bi--stability limit is crossed
at some latitude, a steep increase of the mass flux 
will happen between this latitude and the equator, possibly
leading to a huge equatorial ring.

With the expressions of the flux (2.3) and of 
$\vec{g_\mathrm{tot}}$ (2.10), we get for the mass flux

\begin{eqnarray}
\frac{\Delta\dot{M}(\vartheta)}{\Delta \sigma}
\simeq  A \; \left[\frac{L(P)}{4\pi
GM_\star (P)}\right]^{\frac{1}{\alpha}}
\frac{g_\mathrm{eff} [1+\zeta(\vartheta]^{\frac{1}{\alpha}}}
{(1 - \Gamma_{\Omega}(\vartheta))^{\frac{1}{\alpha}-1}} \nonumber \\[2mm] 
 \quad {\mathrm{with}} \quad
A = \left(k\alpha\right)^{\frac{1}{\alpha}} \left(
\frac{1-\alpha}{\alpha}\right)^{\frac{1-\alpha}{\alpha}} \; ,
\end{eqnarray}

\noindent
with M$_{\star}$ given by Eq. (2.4). 
We  notice the gravity--effect, which favours mass loss 
at the pole, where the total gravity is higher, and the $\kappa$--effect
which favours high mass loss where $\alpha$ is small. 
The proximity to the Eddington limit will 
enhance the mass flux due to the term $\Gamma_{\Omega}(\vartheta)$,
while rotation enhances the mass flux through both the terms M$_{\star}$
and $\Gamma_{\Omega}(\vartheta)$.
In the theory of radiatively driven
winds, the total opacity at a given optical depth is expressed with the 
force multipliers in terms of the electron scattering opacity
$\kappa_\mathrm{es}$. This means that in Eq. (4.22), 
$\Gamma_{\Omega}(\vartheta)$ is just

\begin{eqnarray}
\Gamma_{\Omega}(\vartheta) =
\frac{ \kappa_\mathrm{es} \; L(P)[1+\zeta (\vartheta)]}{4 \pi 
cGM \left( 1 - \frac{\Omega^2}{2 \pi G \rho_{\rm{m}}}  \right) } \; .
\end{eqnarray}

\noindent
The dependence on latitude of $\Gamma_{\Omega}(\vartheta)$
 would only come  through the
term $\zeta(\vartheta)$.

\subsection{Dependence of the global mass loss rates on rotation}

Let us estimate how the global mass loss rates depend on the
rotation velocities and on the proximity of the $\Omega \Gamma$
limit. For that we henceforth neglect the small corrective term  $\zeta(\vartheta)$ in 
the expression of the flux. We have, if $\Sigma(\omega)$ is the 
total surface

\begin{eqnarray}
\frac{\dot{M}}{\Sigma(\omega)} 
\simeq  A \; \left[\frac{L(P)}
{4 \pi G M_{\star}} \right]^{\frac{1}{\alpha}} 
\overline{\left( \frac{g_{\mathrm{eff}}} {(1 - \Gamma_{\Omega})
^{\frac{1}{\alpha}-1}}\right)} \simeq  \nonumber  \\[2mm]
 A \; \left[\frac{L(P)}
{4 \pi G M_{\star}} \right]^{\frac{1}{\alpha}} 
\frac{\overline{g_{\mathrm{eff}}}} {(1 - \Gamma_{\Omega})
^{\frac{1}{\alpha}-1}} \; ,
\end{eqnarray}

\noindent
since $\Gamma_{\Omega}$ is independent on $\vartheta$,
and with appropriate $\alpha$-- and $A$--values.
For the average effective
gravity, we have

\begin{equation}
\overline{g_{\mathrm{eff}}} = \frac{
\int \int \vec{g_{\rm{eff}}} \cdot \vec{d \sigma}}{\Sigma(\omega)}
= \frac{4 \pi G M_{\star}}{\Sigma(\omega)}  \; ,
\end{equation}

\noindent
after integration over the stellar surface which is an isobar
(cf. Appendix). This leads to 
the following expression for the total mass loss rate from the star

\begin{equation}
\dot{M} 
\simeq \frac{ A \; L(P)^{\frac{1}{\alpha}}}
{\left(4 \pi G M \right)^{\frac{1}{\alpha} - 1}
\left[ 1 - \frac{\Omega^2}
{2 \pi G \rho_{\rm{m}}} \right]
^{\frac{1}{\alpha} - 1} (1- \Gamma_{\Omega})^{\frac{1}{\alpha} -1}} \; .
\end{equation}

\noindent
This relation expresses how the total mass loss rate from a star 
depends on mass, luminosity, Eddington factor and rotation,
(see Fig. A1 for a simple expression of the rotation parameter). If we
omit rotation, Eq. (4.26) is identical to the typical relations used
in literature (cf. Pauldrach et al. \cite{Pau86}; Lamers \cite{Lam97}).
 The amplitude of the effects 
very much depends on the opacity and in particular on the value of the force 
multiplier $\alpha$.

Let us consider a rotating star with angular velocity $\Omega$
and a non--rotating star of the same mass $M$ at about the same
location in the HR diagram. 
The ratio of their mass loss rates can be  written,

\begin{equation}
\frac{\dot{M} (\Omega)} {\dot{M} (0)} =
\frac{\left( 1  -\Gamma\right)
^{\frac{1}{\alpha} - 1}}
{\left[ 1 - \frac{\Omega^2}
{2 \pi G \rho_{\rm{m}}} \right]
^{\frac{1}{\alpha} - 1} (1-\Gamma_\mathrm{\Omega})^{\frac{1}{\alpha} -1}}
\; ,
\end{equation}

\noindent
where $\Gamma$ is the Eddington ratio corresponding to electron 
scattering opacity for the non--rotating star. From Eq.
(4.23), we have the relation

\begin{equation} 
\Gamma_{\Omega} = \frac{\Gamma}{1 - \frac{\Omega^2}{ 2 \pi G 
\rho_\mathrm{m}}} \; .
\end{equation}

\noindent
We get finally

\begin{equation}
\frac{\dot{M} (\Omega)} {\dot{M} (0)} =
\frac{\left( 1  -\Gamma\right)
^{\frac{1}{\alpha} - 1}}
{\left[ 1 - \frac{\Omega^2}
{2 \pi G \rho_{\rm{m}}}-\Gamma \right]
^{\frac{1}{\alpha} - 1}}
\; .
\end{equation}

\noindent
If $\Omega = 0 $, this ratio is of course equal to 1.
This ratio, which is the main result of this work, can also be expressed
with the ratio  $v/v_\mathrm{crit,1}$ of the  rotational velocity $v$
to the critical velocity given by the usual Eq. (3.14). 
From Eq. (A7) in the Appendix, we have
$\frac{\Omega^2}{2 \pi G \rho_{\mathrm{m}}} \simeq
\frac{4}{9} \frac{v^2}{v_\mathrm{crit, 1}^2} $ over a large
range of values (cf. Fig. A1), thus one can write

\begin{equation}
\frac{\dot{M} (\Omega)} {\dot{M} (0)} \simeq
\frac{\left( 1  -\Gamma\right)
^{\frac{1}{\alpha} - 1}}
{\left[ 1 - 
\frac{4}{9} (\frac{v}{v_\mathrm{crit, 1}})^2-\Gamma \right]
^{\frac{1}{\alpha} - 1}} \; .
\end{equation}

\noindent
For a star with a small Eddington factor, it simplifies to

\begin{equation}
\frac{\dot{M} (\Omega)} {\dot{M} (0)} \simeq
\frac{1}
{\left[ 1 - 
\frac{4}{9} (\frac{v}{v_\mathrm{crit, 1}})^2 \right]
^{\frac{1}{\alpha} - 1}} \; .
\end{equation}

Equation (4.31) shows that the effects of rotation on the $\dot{M}$--rates
remain moderate in general. This is in agreement with the results by
Owocki et al. (\cite{Ow96}), by Owocki and Gayley (\cite{Ow97}) and
by Glatzel (\cite{Gla98}), and also with more elaborate non--LTE 
1--D and 2--D models by Pauldrach  et al. (\cite{Pau86}), 
Petrenz and Puls (\cite{PP00}). However, this is only true for
stars far enough from the Eddington limit. \textit{When $\Gamma$ is 
significant, rotation may drastically increase the mass loss rates 
as shown by (4.29) or (4.30)}. This is particularly the case for
low values of $\alpha$, i.e. for stars with log T$_\mathrm{eff} \leq$
4.30. In the extreme cases where $\Gamma > 0.639$, a moderate
rotation may  even make the denominator of (4.29) or (4.30) to vanish,
thus leading to extreme mass loss.

\begin{table}
\caption{Values of $\Gamma$ at the end of the MS for various initial
stellar masses, and of the 
ratios  $\dot{M} (\Omega)/ \dot{M} (0)$
of the mass loss rates for a star at break--up rotation to that of
a non--rotating star of the same mass and luminosity
 at $\log T_\mathrm{eff} \geq 4.35 $, at
$\log T_\mathrm{eff}$ = 4.30, 4.00 and 3.90.
The empirical force multipliers $\alpha$ by Lamers et al. (\cite
{Lam95}) are used.} \label{tbl1}
\begin{center}\scriptsize
\begin{tabular}{ccccccc}
$M_{\rm ini}$ &  $\Gamma$
&    $\frac{\dot{M} (\Omega)} {\dot{M} (0)}$
&    $\frac{\dot{M} (\Omega)} {\dot{M} (0)}$
&    $\frac{\dot{M} (\Omega)} {\dot{M} (0)}$

&    $\frac{\dot{M} (\Omega)} {\dot{M} (0)}$ \\


 &  & $\alpha$ = 0.52 & $\alpha$ = 0.24 & $\alpha$ = 0.17
 & $\alpha$ = 0.15 \\
\hline
      &      &         &         &       \\
 120 & 0.903 & $\infty $ & $\infty$  & $\infty $ & $\infty $  \\
 85  & 0.691 & $\infty $ & $\infty$  & $\infty $ & $\infty $  \\

  60 & 0.527 & 3.78  & 96.2  &  1130  & 3526  \\
  40 & 0.356 & 2.14  & 13.6  &  55.3  & 106.0 \\
  25 & 0.214 & 1.76  &  7.02 &  20.1  &  32.6 \\
  20 & 0.156 & 1.67  &  5.87 &  15.2  &  23.6 \\
  15 & 0.097 & 1.60  &  5.04 &  12.1  &  18.1 \\
  12 & 0.063 & 1.57  &  4.68 &  10.8  &  15.8 \\
   9 & 0.034 & 1.54  &  4.41 &   9.8  &  14.2 \\
     &       &       &       &       \\

\hline
\end{tabular}
\end{center}
\end{table}

Table 1 shows some  numerical results based on Eq. (4.29).
For different initial stellar masses in the Geneva models at Z =0.02
(Schaller et al. \cite{Scha92}), the $\Gamma$ factors at the end of the 
Main Sequence (MS) phase are given, as well as the predicted 
ratios  $\dot{M} (\Omega)/ \dot{M} (0)$
of the mass loss rates for a star at break--up rotation and for
a non--rotating star of the same mass and luminosity. These ratios 
are given at $\log T_\mathrm{eff} \geq  4.35 $ ($\alpha = 0.52$),  
at $\log T_\mathrm{eff}$ = 4.30 ($\alpha = 0.24$),
at $\log T_\mathrm{eff}$ = 4.00 ($\alpha = 0.17$) and 
at $\log T_\mathrm{eff}$ = 3.90 ($\alpha = 0.15$)
for the same value of 
$\Gamma$. This  covers the range of
the typical $T_\mathrm{eff}$ of  OB and
LBV stars, the differences with  T$_\mathrm{eff}$ result
from the differences in the $\alpha$--parameter. The indication
$\infty$ in Table 1 means that the bracket term
in (4.29) or (4.30) may vanish at maximum rotation, which leads
to extreme mass outflows.

The ratios $\dot{M} (\Omega)/ \dot{M} (0)$
keep quite moderate even at extreme rotation for MS stars
up to 40 M$_{\odot}$, while for MS stars above 60 M$_{\odot}$ they
can become very large. These ratios may also  be very large for 
B--type supergiants and LBV stars. In particular, we notice that for
stars close to the Humphreys--Davidson limit the ratios
$\dot{M} (\Omega)/ \dot{M} (0)$
may diverge. Such stars are typically at the $\Omega \Gamma$--limit.
For log T$_\mathrm{eff} \leq 4.30$, the force multiplier 
$\alpha$ is also very small, which favours extreme mass loss. On the
whole, it is striking that the domain where 
$\dot{M} (\Omega)/ \dot{M} (0)$
has the possibility to diverge so closely corresponds
to the observed domain of LBV stars. The present results
will enable us to better specify the changes of the 
$\dot{M}$ rates in massive star models.

\section{Conclusion}

We conclude that the concept of an $\Omega \Gamma$--limit reached
during the evolution of the most massive stars is not an artefact,
but the existence of this limit is confirmed by consistent developments 
based on the von Zeipel theorem. However, we emphasize that the expression
currently used for the critical velocity is not correct. We
have also clarified the dependence of the  mass loss rates on 
the rotation velocities in the general case.

We can make the following remarks on the various limits:

\textbf{--1. The $\Gamma$--limit :} The mass
loss rates grow steeply as the Eddington limit is approached, even 
in absence of rotation. This is a well known result of the classical
wind theory.

 \textbf{--2. The $\Omega$--limit :} We see that the case of 
only rotational effects does not apply for O--type stars and even for 
the early B--type stars, since they always  have a significant 
$\Gamma$--value. Only for spectral types later than B3 on
the MS, the $\Gamma$ term can be ignored.
In the framework of the radiative wind theory, the growth of
the mass loss rates remains limited.

\textbf{--3. The  $\Omega \Gamma$--limit :} This general case is met
for rotating OB stars, LBV stars, supergiants and 
Wolf--Rayet  stars,  because both $\Gamma$ 
and rotation are important. The bracket in (4.29)
is reduced by  rotation and  by the proximity to the Eddington
limit. As shown by Table 1,  both effects produce  steep 
enhancements of the mass loss
rates, especially for lower  $T_\mathrm{eff}$ since $\alpha$ is
lower. This may explain the very large mass loss rates for LBV stars, 
blue and yellow supergiants (cf. de Jager et al.
\cite{deJ88}). If the ratio 
$\Gamma = \frac{\kappa_\mathrm{es} L}{4 \pi c GM}$ is bigger than 0.639,
the break--up limit is reached for reduced rotation velocities,
as illustrated by Fig. 1. Then, extremely high mass loss rates may
occur, a situation  likely corresponding to the case of the LBV stars 
and maybe also to some WR stars.

Some words of caution are necessary. The $\dot{M}$--rates given
here are the values predicted in the framework of the
radiative wind theory. It is probable that close to break--up
several other effects not included here may intervene, such as
important horizontal fluxes,  formally
vanishing $T$--  and $P$-- gradients, instabilities, etc... Also, we
may point out that if the flux vanishes, the radiative wind theory
should  not apply. Thus, for the detailed physics of the break--up,
more complex analyses are certainly needed.

Finally, we note that it was
 generally believed that in addition to $L$ and
$T_\mathrm{eff}$, the mass loss rates only depend on
metallicity Z. We see here another dependence which is
quite significant and may introduce some scatter in the
values of the $\dot{M}$--rates. Thus, we may expect 
that for a given initial mass  the evolution is very different
according to rotation, due to both rotational mixing,
meridional circulation (Maeder and Zahn \cite{MZ98}) and to 
the induced differences in the mass loss rates.

\appendix
\section {The equation of the stellar surface in a rotating 
star with high radiation pressure}

Shellular rotation, with an angular velocity  $\Omega$ constant  on 
horizontal surfaces, was  proposed by Zahn (\cite{Za92}). This
rotation law results from strong horizontal geostrophic--like turbulence 
which homogeneizes rotation on the horizontal surfaces. As noted 
by Meynet and Maeder (\cite{MM97}), the isobars for shellular rotation
are identical to the equipotentials  of the conservative case, which
are 

\begin{equation}
\Psi = \frac{GM}{r(\vartheta)} + \frac{1}{2} \Omega^2 r^2 (\vartheta)
\sin^2 \vartheta = \mathrm{const}
\end{equation}

\noindent
The components of the effective gravity are

\begin{eqnarray}
g_{\mathrm{eff}, r} & = & \frac{\partial\Phi}{\partial r} +
 \Omega^2 r \sin \vartheta \nonumber \\[2mm] 
g_{{\mathrm{eff}}, \vartheta} & = & \frac{1}{r} 
\frac{\partial\Phi}{\partial\vartheta}
+ \Omega^2 r \sin\vartheta\cos\vartheta
\end{eqnarray}

\noindent
where $\Phi = GM/r$. In vectorial form, one can write

\begin{equation}
\vec{\nabla} P= -\rho \vec{g_\mathrm{eff}} = -\rho(\vec{\nabla}\Psi 
- r^2 \sin^2 \vartheta  \; \Omega \vec{\nabla}\Omega)  \; .
\end{equation}

\noindent
This equation is interesting: it shows that if $\Omega$ is
 constant on isobars, $\Psi$ is also constant on isobars. Moreover,
for a motion $d \vec{s}$,
the equation of the surface must satisfy 

\begin{equation}
\vec{g_\mathrm{tot}} \cdot \vec{ds} = 0\; .
\end{equation}

\noindent Since we have 

\begin{equation}
\vec{g_\mathrm{tot}} = \vec{g_\mathrm{eff}}
\left[ 1 - \Gamma_{\Omega} \right] \; ,
\end{equation}

 \begin{figure}[tb]
  \resizebox{\hsize}{!}{\includegraphics{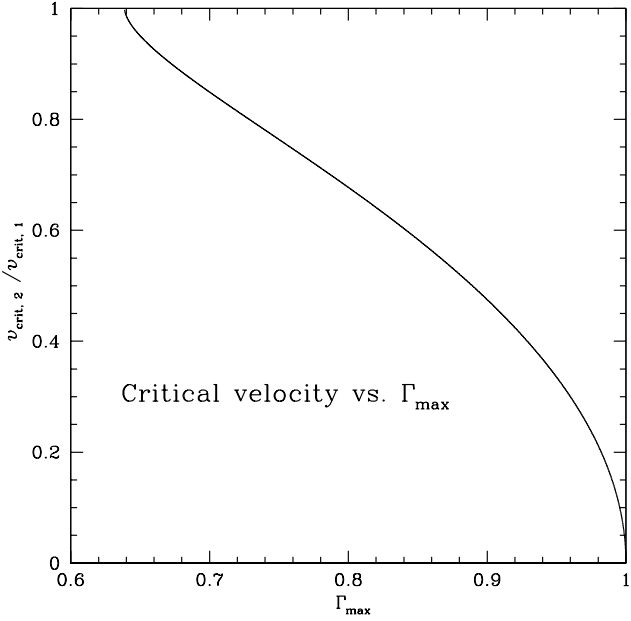}}
  \caption{ The rotation parameter
$\frac{\Omega^2}{2 \pi G \rho_{\rm{m}}}$ as a function of the 
square of the ratio of the rotational velocity to the critical 
velocity (3.14).  The maximum value of
$\frac{\Omega^2}{2 \pi G \rho_{\mathrm{m}}}$ = 0.361, as seen above.}
\label{rhomv2}
\end{figure}

\noindent
this implies that the surface is perpendicular to 
$\vec{g_\mathrm{tot}}$, thus to 
$\vec{g_\mathrm{eff}}$ and  to the P---gradient. The surface is 
an isobar, also if the radiation pressure is important.
The equation of the surface can therefore be represented
by Eq. (A1) and the procedure to calculate is quite conventional,
i.e.

\begin{equation}
\frac{1}{x} + \frac{4}{27} \omega^2 x^2 \sin^2 \vartheta = 1
\end{equation}

\noindent
with $ x = \frac{R}{R_\mathrm{p}}$. At break--up, the equatorial
radius $R_\mathrm{eb}$ equals 1.5 times the polar radius $R_\mathrm{pb}$.

In the above demonstration, there is no need of the assumption
$\Omega = \Omega(r)$, we just need the assumption that $\Omega$
is constant on horizontal surface, which is less restrictive.
This is true, whether the radiative acceleration is important
or not.

The critical rotation parameter naturally appearing in this work
was the ratio $\frac{\Omega^2}{2 \pi G \rho_{\rm{m}}}$. Account
must be given to the change of the average density of the star
with rotation. We can  express this ratio by (3.16) or
in term of the actual rotational velocity $v$ 
and of the critical velocity $v_\mathrm{crit, 1}$ (3.14), 

\begin{equation}
\frac{\Omega^2}{2 \pi G \rho_{\mathrm{m}}}=
\frac{4}{9} \frac{v^2}{v_\mathrm{crit, 1}^2} V^{\prime}(\omega)
\frac{R^2_{\mathrm{pb}}}{R^2_{\mathrm{e}}(\omega)} \; .
\end{equation}

\noindent
The relation between $\frac{\Omega^2}{2 \pi G \rho_{\rm{m}}}$
and the ratio $\frac{v^2}{v_\mathrm{crit, 1}^2}$ is illustrated
in Fig. A1.
The  product  $V^{\prime}(\omega)
\frac{R^2_{\mathrm{pb}}}{R^2_{\mathrm{e}}(\omega)}$  has a limited
range of variation, being  equal to 1 for zero
rotation and to 0.813 at break--up velocity. This means that for a 
crude estimate at low or moderate velocities, one may just ignore 
this product in Eq. (A7).

\begin{acknowledgements}
We express our  thanks to Dr. Joachim Puls for very valuable remarks
during this work.
\end{acknowledgements}

\end{document}